\documentclass[prb,aps,superscriptaddress,
                           noshowpacs,twocolumn,reprint,noshowkeys,
                           amssymb,amsmath,a4paper]{revtex4-1}

\usepackage{microtype} 
\usepackage{xspace} 

\usepackage{txfonts}  
\usepackage{txfontsb} 

\usepackage{bm} 

\usepackage{xcolor}
\usepackage{graphicx}

\usepackage{hyperref}
\hypersetup{colorlinks,
            linkcolor={blue!75!black!80!yellow},
            citecolor={blue!75!black!80!yellow},
                urlcolor={blue!75!black!80!yellow}  }

\usepackage{placeins}

\usepackage{siunitx}
\usepackage[centering,hmargin=16mm,tmargin=30mm,bmargin=26mm]{geometry}

\newcommand{\SSS}{\scriptscriptstyle}
\newcommand{\DS}{\displaystyle}

\newcommand{\Ee}{{\rm e}}
\newcommand{\Dd}{{\rm d}}

\newcommand{\Ii}{{\rm i}}

\newcommand{\Ez}{\hat{\mathbf{e}}_{z}}
\newcommand{\rv}{\mathbf{r}}
\newcommand{\jv}{\mathbf{j}}

\DeclareMathOperator{\sgn}{sgn}

\newcommand{\OmegaC}{\omega_{\text{c}}}

\newcommand{\EpsA}{\epsilon_{{\SSS\text{A}}}}
\newcommand{\EpsB}{\epsilon_{{\SSS\text{B}}}}

\newcommand{\DdA}{d_{\SSS\text{A}}}
\newcommand{\DdB}{d_{\SSS\text{B}}}

\newcommand{\JR}{j_{\SSS\text{R}}}
\newcommand{\JL}{j_{\SSS\text{L}}}
\newcommand{\JD}{j_{\SSS\text{D}}}

\newcommand{\VG}{V_{\text{g}}}
\newcommand{\BBM}{\bar{B}_{\text{m}}}
\newcommand{\DBM}{\Delta B_{\text{m}}}

\newcommand{\ie}{i.e.\@\xspace} 

\usepackage[textsize=tiny,
backgroundcolor=black!8!white,
bordercolor=black!10!white,
linecolor=black!60!white]{todonotes}

\makeatletter
\renewcommand\@make@capt@title[2]{%
        \@ifx@empty\float@link{\@firstofone}{\expandafter\href\expandafter{\float@link}}%
        \sffamily{\textbf{#1}}\@caption@fignum@sep#2
}%

\makeatother

\usepackage{titlesec}
\titleformat{\paragraph}[runin]{\normalfont\normalsize\bfseries}{\theparagraph}{1em}{}
\titlespacing*{\paragraph}{0pt}{1.25ex plus 1ex minus .2ex}{1em}

\begin{document}

\title{Magnetically-defined topological edge plasmons in edgeless electron gas}

\author{Dafei Jin}\email{D. J. and Y. X. contributed equally to this work.}
\author{Yang Xia}\email{D. J. and Y. X. contributed equally to this work.}
\affiliation{Nanoscale Science and Engineering Center, University of California, Berkeley, California 94706, USA}
\author{Thomas Christensen}
\affiliation{Department of Physics, Massachusetts Institute of Technology, Cambridge, Massachusetts 02139, USA}
\author{Siqi Wang}
\author{King Yan Fong}
\affiliation{Nanoscale Science and Engineering Center, University of California, Berkeley, California 94706, USA}
\author{Matthew Freeman}
\affiliation{National High Magnetic Field Laboratory, Tallahassee, Florida 32310, USA}
\author{Geoffrey C. Gardner}
\affiliation{Department of Physics and Astronomy and Birck Nanotechnology Center, Purdue University, West Lafayette, Indiana 47907, USA}
\affiliation{Station Q Purdue and School of Materials Engineering, Purdue University, West Lafayette, Indiana 47907, USA}
\author{Saeed Fallahi}
\affiliation{Department of Physics and Astronomy and Birck Nanotechnology Center, Purdue University, West Lafayette, Indiana 47907, USA}
\author{Qing Hu}
\affiliation{Department of Mechanical Engineering, Massachusetts Institute of Technology, Cambridge, Massachusetts 02139, USA}
\author{Yuan Wang}
\affiliation{Nanoscale Science and Engineering Center, University of California, Berkeley, California 94706, USA}
\author{Lloyd Engel}
\affiliation{National High Magnetic Field Laboratory, Tallahassee, Florida 32310, USA}
\author{Michael J. Manfra}
\affiliation{Department of Physics and Astronomy and Birck Nanotechnology Center, Purdue University, West Lafayette, Indiana 47907, USA}
\affiliation{Station Q Purdue and School of Materials Engineering, Purdue University, West Lafayette, Indiana 47907, USA}
\affiliation{School of Electrical and Computer Engineering, Purdue University, West Lafayette, Indiana 47907, USA}
\author{Nicolas X. Fang}
\affiliation{Department of Mechanical Engineering, Massachusetts Institute of Technology, Cambridge, Massachusetts 02139, USA}
\author{Xiang Zhang}\email{Corresponding author. Email: xiang@berkeley.edu}
\affiliation{Nanoscale Science and Engineering Center, University of California, Berkeley, California 94706, USA}



\date{\today}

\maketitle
\pretolerance=8000 


{\bfseries
Topological materials bear gapped excitations in bulk yet protected gapless excitations at boundaries~\cite{Qi2011RMP,Lu2014NatPhoton}. Magnetoplasmons (MPs), as high-frequency density excitations of two-dimensional electron gas (2DEG) in a perpendicular magnetic field~\cite{Ando1982RMP,Kushwaha2001SSR}, embody a prototype of band topology for bosons~\cite{Jin:2016,Jin2017PRL}. The time-reversal-breaking magnetic field opens a topological gap for bulk MPs up to the cyclotron frequency~\cite{Zudov2003PRL,Gao2016NatCommun}; topologically-protected edge magnetoplasmons (EMPs) bridge the bulk gap and propagate unidirectionally along system's boundaries~\cite{Mast1985PRL,Glattli1985PRL,Fetter:1986,Volkov1988JETP}.
However, all the EMPs known to date adhere to physical edges where the electron density terminates abruptly~\cite{Ashoori1992PRB,Balev1997PRB,Kumada2014PRL}. This restriction has made device application extremely difficult. Here we demonstrate a new class of topological edge plasmons -- domain-boundary magnetoplasmons (DBMPs), within a uniform edgeless 2DEG. Such DBMPs arise at the domain boundaries of an engineered sign-changing magnetic field and are protected by the difference of gap Chern numbers ($\pm1$) across the magnetic domains. They propagate unidirectionally along the domain boundaries and are immune to domain defects~\cite{Jin:2016}. Moreover, they exhibit wide tunability in the microwave frequency range under an applied magnetic field or gate voltage. Our study opens a new direction to realize high-speed reconfigurable topological devices~\cite{Mahoney2017PRX,Fang2012NatPhoton,Bahari2017Science}.
}

In this work, we present the first experimental observation of a new class of topological edge plasmons, domain-boundary magnetoplasmons (DBMPs), at microwave frequencies in a high-mobility GaAs/AlGaAs heterojunction. In contrast to the traditional wisdom, where edge magentoplasmons (EMPs) must rely on a space-varying electron density $n(\rv)$, in our scenario, the DBMPs are defined by a space-varying magnetic field $B(\rv)=B(\rv)\Ez$ embedded into a uniform 2DEG~\cite{Ye1995PRL,Nogaret2000PRL,Reijniers2000JPCM,Yasuda2017Science}. A custom-shaped NdFeB strong permanent magnet, placed immediately above the heterojunction, produces a sign-changing magnetic field around $0.15$~T in magnitude, sufficient to gap bulk MPs in each magnetic domain. The $10^7$~cm$^2$~V$^{-1}$~s$^{-1}$ high electron mobility in this system affords an ultra-long relaxation time of hundreds of picoseconds and ultra-low damping rate of only a few gigahertz, superior to any other existing 2DEG systems \cite{Mast1985PRL,Bolotin2008PRL,Ohtomo2004Nature}. By measuring microwave resonant spectra, we clearly verify the existence and nonreciprocal nature of DBMPs. Their excitation frequencies display a unique dependence on both an applied magnetic field and gate voltage, differing substantially from the conventional EMPs in several intriguing aspects. Our theoretical prediction and experimental observation show excellent mutual agreement.

\begin{figure*}[htb]
\centerline{\includegraphics[scale=0.7]{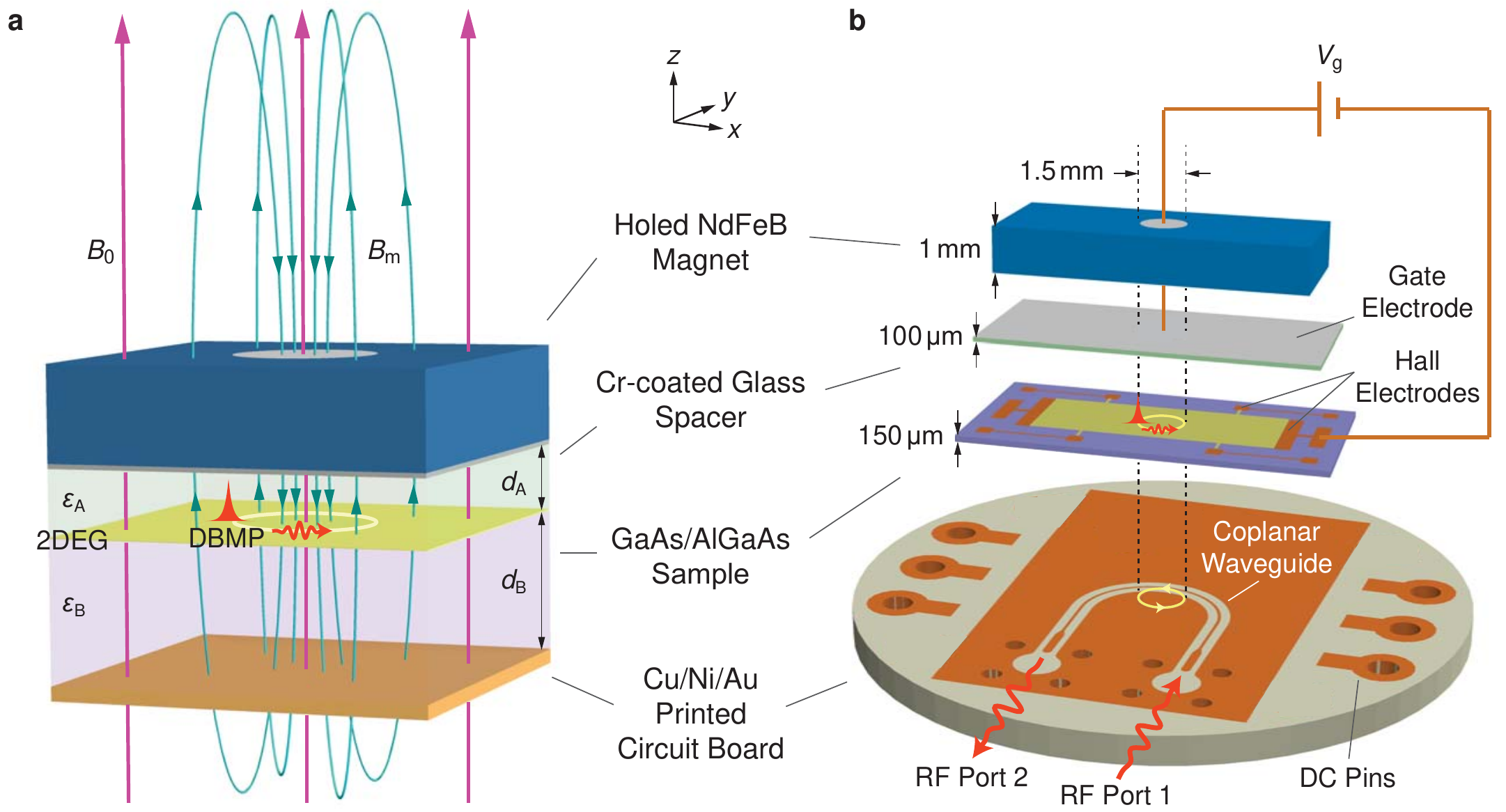}}
\caption{%
        \textbf{Magnetoplasmonic device.}
        \textbf{a}, Conceptual layout.
        \textbf{b}, Device design and PCB layout. The 2DEG is formed at the interface of a GaAs/AlGaAs heterojunction, cladded between a glass spacer and the substrate, and finally enclosed in a metallic cavity. A holed NdFeB magnet on the top provides an oppositely-signed magnetic field at the 2DEG to permit domain-boundary magnetoplasmons (DBMPs)  traveling unidirectionally along the magnetic-domain boundary. Microwaves transmitted along the coplanar waveguide on a printed circuit board excite the DBMPs. An applied uniform magnetic field $B_0$ or a gate voltage $V_{\text{g}}$ can tune the DBMPs.
        \label{fig:device}}
\end{figure*}

\noindent{\bf System design}
Figure~\ref{fig:device} illustrates the layout of our magnetoplasmonic device.
Conceptually (Fig.~\ref{fig:device}a), a 2DEG in a GaAs/AlGaAs heterojunction (see Methods) is cladded above and below by a fused silica (glass) spacer and a GaAs substrate, respectively, of thicknesses $\DdA = 100~\mu$m and $\DdB = 150~\mu$m, and permittivities $\EpsA=3.8$ and $\EpsB=12.8$. This dielectric-2DEG-dielectric structure is enclosed in a metallic cavity along $z$, terminating at the spacer's top and substrate's bottom. A holed NdFeB permanent magnet, atop the upper cavity wall, projects a circular magnetic field $\mathbf{B}_{\text{m}}(\rv) = B_{\text{m}}(r)\Ez$ onto the 2DEG. The sign of $B_{\text{m}}(r)$ changes abruptly across the projection of the hole's radius, $a = 0.75~$mm, producing adjacent oppositely-signed magnetic domains (see Methods). The entire 2DEG is additionally exposed to a tunable homogeneous magnetic field $\mathbf{B}_0(\rv) = B_0\Ez$ from a superconducting coil, allowing an overall shift of the magnetic field profile.

In practice (Fig.~\ref{fig:device}b), the heterojunction sample has a 12~mm~$\times$~6~mm rectangular footprint. A 9~mm~$\times$~3~mm Hall bar is fabricated atop of it, allowing \textit{in situ} measurements and control of the 2DEG electron concentration $n_0$. The fused silica spacer is topped by a 100~nm thick e-beam evaporated Cr-coating, serving simultaneously as upper cavity wall and gate electrode \cite{Hatke2015NatCommun,Mi2017arXiv}. A gate voltage of $V_\text{g} \sim \pm 100~$V can be applied across the Cr-coating--Hall bar junction to tune the electron concentration. The sample-spacer-magnet assembly is glued by Poly(methyl methacrylate) (PMMA) onto a customized Cu printed circuit board (PCB) with a 5~$\mu$m Ni and 200~nm Au surface finish. The PCB hosts a coplanar waveguide (CPW) connecting RF Ports 1 and 2 with mini-SMP connectors \cite{Hatke2015NatCommun}. By design, the CPW has a 50~$\Omega$ impedance with the sample-magnet assembly loaded. The CPW signal line is aligned tangentially to the projected circle from the hole of magnet so as to maximize the microwave-DBMP coupling.


\noindent{\bf Theoretical prediction}
The main physics of MP system can be captured by the continuity equation and a constitutive equation containing the longitudinal Coulomb force and transverse Lorentz force:
\begin{subequations}\label{eqs:governing}
                \begin{align}
                &\omega\rho(\rv,\omega) =  -\Ii \nabla\cdot \jv(\rv,\omega),\\
                &\omega\jv(\rv,\omega) = -\Ii \frac{e^2}{m_*}n(\rv)\nabla\Phi(\rv,\omega) - \OmegaC(\rv)\jv(\rv,\omega)\times\hat{\mathbf{e}}_{z}.
        \end{align}
\end{subequations}
Here, $\jv$ and $\rho$ are the surface current and charge densities, evaluated at frequencies $\omega$ and in-plane positions $\rv$. $\Phi(\rv,\omega) = \int V(\rv-\rv')\rho(\rv',\omega)\,\Dd^2{\rv'}$ is the self-consistent potential due to the (screened) Coulomb interaction $V$. $\OmegaC(\rv)=eB(\rv)/cm_*$ is a space-varying cyclotron frequency, with $m_*$ the electron effective mass. As elaborated below, even with a constant electron density $n(\rv)=n_0$, topologically-protected DBMPs can reside at boundaries of sign-changing magnetic domains solely defined by the spatial profile $B(\rv)$ and $\OmegaC(\rv)$~\cite{Jin:2016}.

The total magnetic field, $B(r)  = B_0 + B_{\text{m}}(r)$, is the sum of a tunable, uniform field $B_0$ from the superconducting coil, and a fixed, $r$-dependent field $B_{\text{m}}(r)$ from the holed NdFeB permanent magnet. The latter is well-approximated by a step function,
\begin{equation}\label{eq:Bm}
B_{\text{m}}(r) \simeq \BBM + \sgn(r-a)\DBM.
\end{equation}
Here, $\DBM$ contributes an equal-magnitude sign-changing jump at $r=a\approx 0.75$~mm, while $\BBM$ accounts for a small, overall shift due to the small distance between magnet and 2DEG. By a combination of finite-element simulations and room-temperature Hall-probe measurements on the surface of magnet, we infer the low-temperature values of each as $\DBM\approx0.14$~T and $\BBM\approx0.01$~T (see Methods).

The presence of cladding dielectrics and encapsulating metals in this system significantly influences the Coulomb interaction and frequency scale of the problem. In momentum space, the Coulomb interaction takes a screened form,
\begin{equation}
V(q) = \frac{2\pi}{q} \beta(q) = \frac{2\pi}{q} \frac{2}{\EpsA \coth(q \DdA)+\EpsB \coth(q \DdB)}  , \label{eq:coulomb}
\end{equation}
with $\beta(q)$ being the $q$-dependent screening function \cite{Fetter:1986,Volkov1988JETP}. The scalar potential and surface charge density are related by $\Phi(q)=V(q)\rho(q)$. The eigenmodes consistent with Eqs.~\eqref{eqs:governing} are eigenstates of a $3\times3$ Hamiltonian $\boldsymbol{\mathcal{\hat{H}}}$ with operator elements~\cite{Jin:2016,Jin2017PRL}. In the circularly symmetric ``potential'' of Eq.~\eqref{eq:Bm}, the eigenmodes decompose according to $\mathbf{R}_{m}(r)\Ee^{\Ii m\varphi}$ with azimuthal angle $\varphi$ and angular wavenumber $m\in\mathbb{Z}$. The radial function $\mathbf{R}_{m}(r)$ can be expanded by the Bessel functions with radial wavenumbers $q_{mn}$, $n\in\mathbb{Z}^+$, which enter the Coulomb interaction Eq.~(\ref{eq:coulomb}) (see Methods).

\begin{figure*}[htb]
\centerline{\includegraphics[scale=0.6]{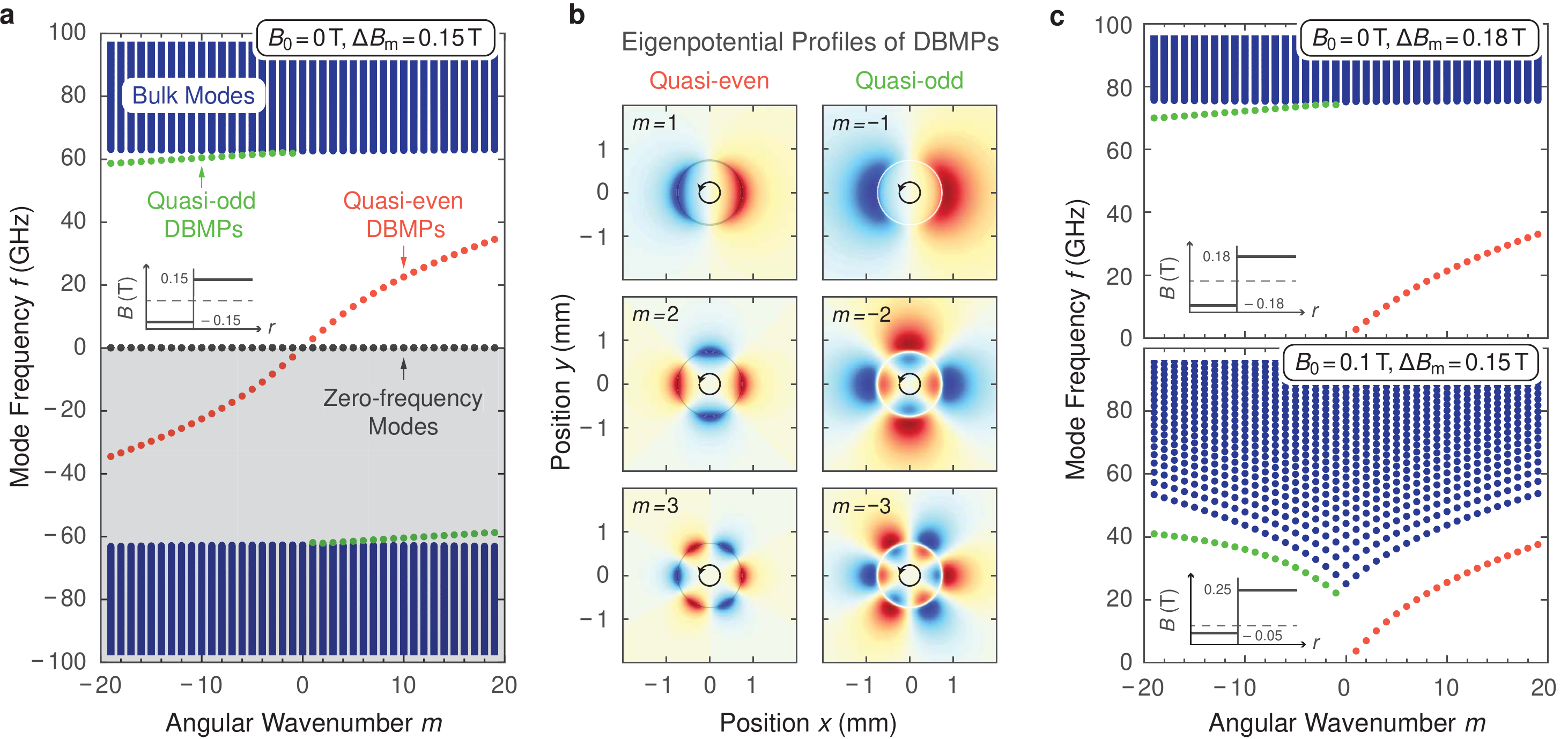}}
\caption{%
        \textbf{Theoretical prediction of DBMPs.}
        \textbf{a}, Magnetoplasmonic dispersion of bulk and edge modes with angular wavenumber $m$ (external field $B_0=0$~T,  material-induced field $\DBM=0.15$~T).
        \textbf{b}, Eigenpotential profiles of quasi-even and quasi-odd DBMPs for $|m|=\text{1, 2, and 3}$ ($B_0$ and $\DBM$ as in a).
        \textbf{c}, Magnetoplasmonic dispersion for $B_0=0$~T, $\DBM=0.18$~T and $B_0=0.1$~T, $\DBM=0.15$~T. Insets indicate the total magnetic field profile $B(r)$ (vertical gray line, $r=a$).
        $n_0=1\times 10^{11}$~cm$^{-2}$ and $\BBM = 0$~T in all panels.
        \label{fig:theory}
        }
\end{figure*}

The resulting plasmonic properties are explored in Fig.~\ref{fig:theory}.
Figure~\ref{fig:theory}a illustrates the magnetoplasmonic dispersion of bulk MP and DBMP modes for $n_0=1\times 10^{11}$~cm$^{-2}$, $B_0=\BBM=0$~T, and $\DBM=0.15$~T. The spectrum exhibits particle-hole symmetry, \ie $\omega_{nm} = -\omega_{n,-m}$, with a zero-frequency band describing static modes~\cite{Jin:2016}.
The bulk MPs in each magnetic domain exhibits a gap from zero frequency to approximately $|\OmegaC(r)|=e|B(r)|/m_*c$.
The band topology of each domain, considered as an extended bulk, is characterized by a topological invariant, the Chern number, equaling $C = -\sgn B(r) = \sgn(a-r)=\pm 1$~\cite{Jin:2016}. The associated gap Chern number $\bar{C}$, also equaling $\pm 1$ in this case, can be identified, whose difference across the domains, $\Delta\bar{C} = 2$, dictates the existence of two unidirectional edge states localized at $r=a$. These conclusions are clearly manifest in Fig.~\ref{fig:theory}a and \ref{fig:theory}b from the existence of quasi-even and quasi-odd DBMP branches (so named due to their asymptotic association with the even and odd DBMPs of a linear domain boundary). Both are unidirectional and exhibit increasing localization with incrementing angular wavenumbers $m$.
We emphasize that these DBMPs drastically differ from the conventional EMPs by being solely magnetically-engineered. Interestingly, they are topologically equivalent to the equatorial Kelvin and Yanai waves of the ocean and atmosphere (with a Coriolis parameter in place of $B(r)$)~\cite{Delplace:2017}.

Figure~\ref{fig:theory}c investigates the dispersion for increased $\DBM$ (from 0.15~T to 0.18~T) and $B_0$ (from 0 to 0.1~T). Comparing to Fig.~\ref{fig:theory}a, increasing $\DBM$ widens the bandgap and decreases the frequencies of the quasi-even DBMPs. Conversely, increasing $B_0$ (but maintaining $B_0<\DBM$) reduces the overall gap---since the cyclotron frequency is lowered in the inner domain---and increases the excitation frequencies of the quasi-even DBMP. This latter behavior further distinguishes our new DBMPs from the traditional EMPs which shift in the opposite direction with increasing $B_0$~\cite{Fetter:1986,Mast1985PRL}. The quasi-odd DBMP branch in Fig.~\ref{fig:theory}c appears non-gapless and hence non-topological in the considered range of $m$: this, however, is remedied at larger $|m|$ where the dispersion bends downwards (not shown), instating an asymptotically gapless behavior  in deference to the topological requirements.

\begin{figure*}[hbt]
        \centerline{\includegraphics[scale=0.68]{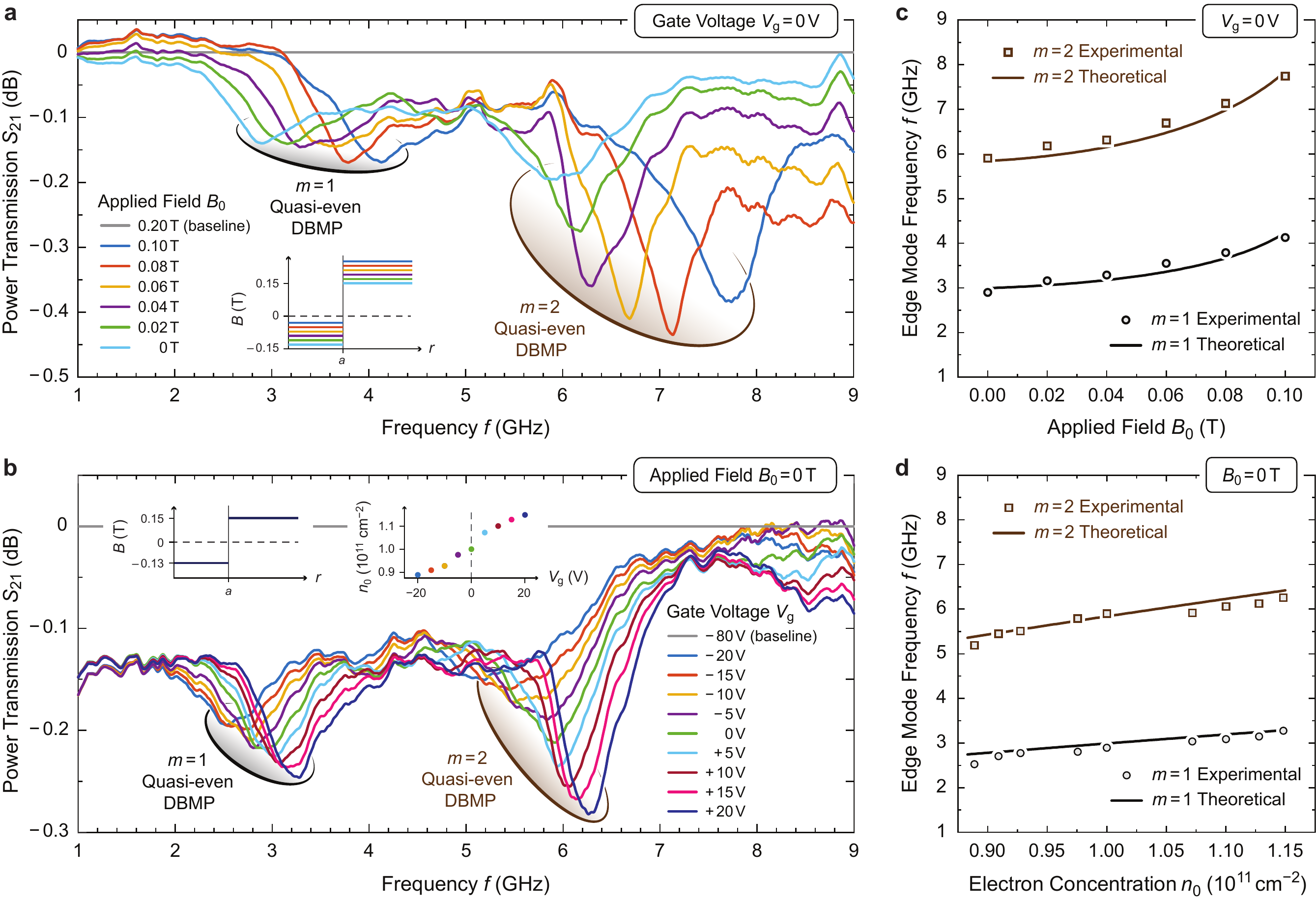}}
        \caption{%
                \textbf{Experimental observation of DBMPs.}
                \textbf{a--b}, Measured power transmission $S_{21}$ (normalized to indicated baselines) as a function of frequency for
                \textbf{a}, varying applied uniform fields $B_0$ ($\VG = \text{0~V}$) and
                \textbf{b}, varying gate voltages $\VG$ ($B_0 = \text{0~T}$). The characteristic absorption dips correspond to quasi-even DBMPs with angular wavenumber $m=\text{1}$ and 2, as indicated. Insets depict the (idealized) total magnetic field profile $B(r)$ across the domains as well as the measured dependence of the electron concentration $n_0$ with $\VG$ (from independent Hall transport measurements).
                \textbf{c--d}, Comparison between experimental observations and theoretical calculations. Dependence of DBMP frequencies with
                \textbf{c}, applied field $B_0$ ($\VG =\text{0~V}$, $n_0 = \text{1}\times\text{10}^{\text{11}}$~cm$^{-\text{2}}$) and
                \textbf{d}, electron concentration $n_0$ ($B_0=\text{0~T}$).
                \label{fig:measurements}
                }
\end{figure*}

\noindent{\bf Experimental observation}
We next seek experimental evidence for the theoretically predicted DBMPs. The device is inserted into a He-3 cryostat running at 0.5~K. An Agilent E5071C Network Analyzer (NA) is used to acquire power transmission $S_{21}$ (Port 1 to Port 2) and $S_{12}$ (Port 2 to Port 1) in the frequency range 300~kHz to 20~GHz~\cite{Hatke2015NatCommun,Mi2017arXiv,Mahoney2017PRX}. In practice, however, our focused frequency range is limited to 1 to 10~GHz, beyond which the cables and NA bear too high loss and noise, prohibiting acquisition of clear signals.
Referring to Figs.~\ref{fig:theory}a and \ref{fig:theory}c, we consequently expect to observe characteristic absorption associated with only the $m=1$ and $2$ quasi-even DBMPs.

In the first series of measurements, we keep the gate grounded, $\VG = 0$~V, and investigate the influence of the applied magnetic field $B_0$ on the resonant absorption of quasi-even DBMPs in $S_{21}$ (Fig.~\ref{fig:measurements}a).
All signals are divided by a reference (denoted baseline). Here, we choose $B_0=0.2$~T as baseline, which provides a high suppression of unwanted low-frequency bulk modes, without exerting too great a torque on the magnet--sample assembly.
For every $S_{21}$-spectrum in Fig.~\ref{fig:measurements}a, each reflecting a single applied field in the range $B_0=0$ to $0.1$~T, we observe two well-defined absorptive resonances, corresponding to the $m=1\text{ and }2$ right-circulating quasi-even DBMPs.
Spanning frequencies from 3 to 4~GHz and 6 to 8~GHz, they exhibit linewidths of approximately 1 to 2~GHz, roughly consistent with the Hall-probe inferred DC damping rate $\gamma\sim2.6$~GHz.
Figure~\ref{fig:measurements}c compares the measured and theoretically predicted resonance frequencies.
First, we observe the excellent mutual agreement in the absence of fitting parameters. Second, we emphasize the monotonously increasing excitation frequencies with increasing $B_0$, which unambiguously differentiates our magnetically-defined DBMPs from the conventional EMPs.

In the second series of measurements, we fix the applied magnetic field $B_0=0$~T, and explore the DBMPs' dependence on the gate voltage $\VG$ (Fig.~\ref{fig:measurements}b).
The baseline is chosen at $\VG=-80$~V, which corresponds to an essentially electron-depleted 2DEG supporting no plasmonic modes.
Once more, every spectrum in Fig.~\ref{fig:measurements}b, each now corresponding to distinct gate voltages in the range $\VG=-20$ to $+20$~V, exhibits two clear absorptive resonances associated with the $m=1\text{ and }2$ quasi-even DBMPs.
Increasing the gate voltage (or, equivalently, the electron concentration $n_0$) increases the DBMP frequency, as expected. Moreover, the extinction depth of each resonance also increases with the $\VG$.
This is consistent with the $f$-sum rule~\cite{YangKall:2015} which dictates a linear increase of integrated extinction with increased $n_0$ (disregarding the negligible spectral dispersion in the microwave-DBMP coupling). Comparing theoretical and experimental observations, in Fig.~\ref{fig:measurements}d, we once again find excellent agreement.

\begin{figure}[htb]
\centerline{\includegraphics[scale=0.68]{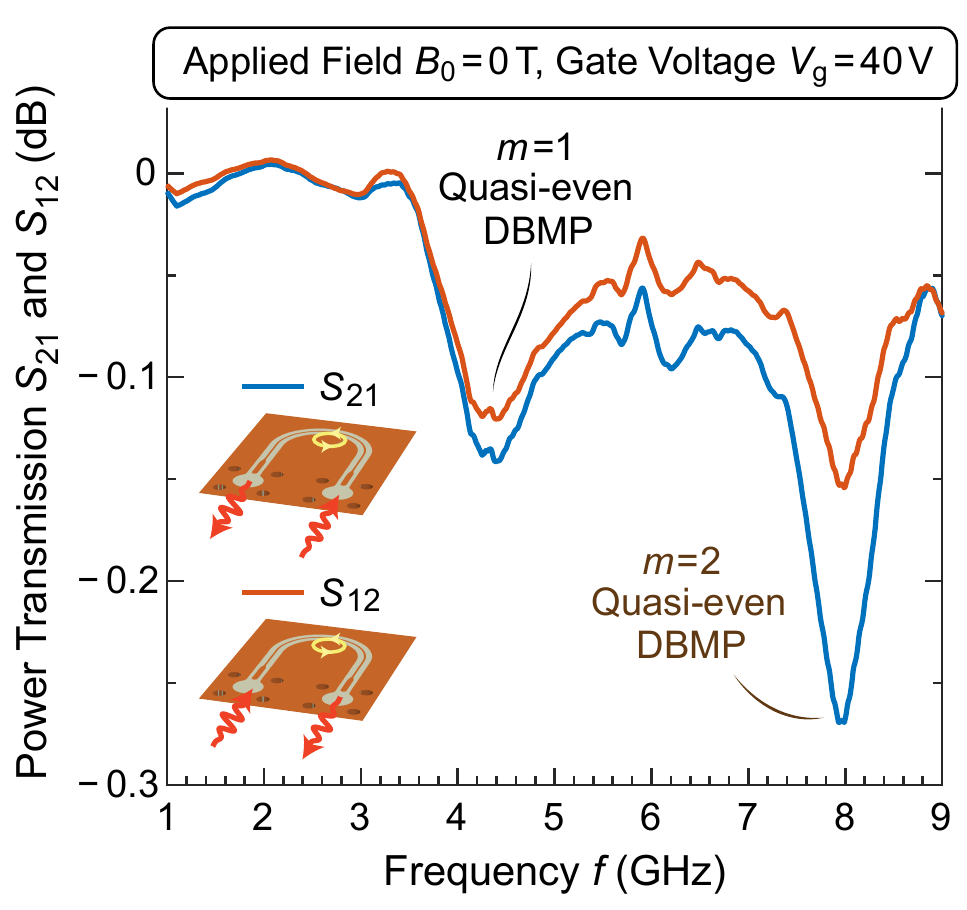}}
\caption{%
        \textbf{Nonreciprocal transmission near DBMPs.}
        Measured power transmission along $S_{21}$ (``easy-coupling'') and $S_{12}$ (``hard-coupling'') directions, respectively, at $B_0 = 0$~T and gate voltage $\VG = 40$~V. The distinct absorption depths manifests the nonreciprocal nature of the $m=\text{1}$ and 2 quasi-even DBMPs.
        \label{fig:isolation}
        }
\end{figure}

Finally, in Fig.~\ref{fig:isolation}, we examine the nonreciprocal properties of the DBMPs in order to explicitly demonstrate the underlying unidirectional character of the DBMPs.
Since the DBMPs are right-circulating in the bandgap (Fig.~\ref{fig:theory}), $S_{21}$ and $S_{12}$ correspond to the ``easy-coupling'' and ``hard-coupling" directions, respectively, of our device (Fig.~\ref{fig:device}b).
Each coupling direction is normalized separately, with baselines taken at $B_0=0.2$~T.
The 2DEG is gated by $\VG=40$~V, ensuring a pronounced extinction depth, and the applied magnetic field is turned off $B_0=0$~T.
In this configuration, the $m=1$ and $2$ quasi-even DBMPs exist at 4.2~GHz and 8.0~GHz, respectively. Comparing $S_{21}$ and $S_{12}$ we observe distinct asymmetry of extinction depth at each resonance, with $S_{12}$ exhibiting shallower extinction. This asymmetry is indicative of the unidirectional character of the DBMPs. The observed isolation ratio $S_{21}/S_{12} = (S_{21}-S_{12})|_\text{dB}$ is small because of the wavelength mismatch between the microwaves in CPW and the DBMPs along the circle. This is mainly a limitation from the CPW evanescent-coupling technique. A fuller assessment of the isolation capabilities could more naturally be enabled by point-source excitation~\cite{Ashoori1992PRB,Wang2009Nature,Fei:2011,Kumada2014PRL}.

\noindent{\bf Conclusion}
We have for the first time realized a new class of topologically-protected edge plasmons, domain-boundary magnetoplasmons, embedded in an edgeless 2DEG. They situate at magnetically-defined domain boundaries, and are topologically distinct from the conventional edge magnetoplasmons. We have experimentally observed and characterized these new DBMPs at microwave frequencies in a high-mobility GaAs/AlGaAs heterojunction under a custom-shaped NdFeB permanent magnet. Our experimental results show remarkable agreement with theoretical calculations. The demonstrated DBMP architecture, if packed with denser magnetic patterns, can be extended to higher frequencies and finer scales, and shall pave the way towards high-speed reconfigurable topological devices~\cite{Mahoney2017PRX,Fang2012NatPhoton,Bahari2017Science}.

\noindent{\bf Acknowledgements}
D.J., Y.X., S.W., K.Y.F., Y.W. and X.Z. acknowledge support from AFOSR MURI (Grant No. FA9550-12-1-0488) and Office of Sponsored Research (OSR) (Award No. OSR-2016-CRG5-2950-03). T.C. acknowledges support from the Danish Council for Independent Research (Grant No. DFF--6108-00667). The National High Magnetic Field Laboratory (NHMFL) is supported by NSF Cooperative Agreement (No. DMR-0654118), the State of Florida, and the DOE. M.F. and L.E., and Microwave Spectroscopy Facility are supported by the DOE (Grant No. DE-FG02-05-ER46212). G.C.G. and M.J.M. acknowledge support from the DOE Office of Basic Energy Sciences, Division of Materials Sciences and Engineering (Award No. {DE{-}SC0006671}), the W. M. Keck Foundation, and Microsoft Station Q. Q.H. and N.X.F. are supported by AFOSR MURI (Grant No. FA9550-12-1-0488).


\bibliographystyle{apsrev4-1}
\bibliography{References}

\clearpage

\noindent{\bf Methods}

\begin{figure}[hb]
\centerline{\includegraphics[scale=0.68]{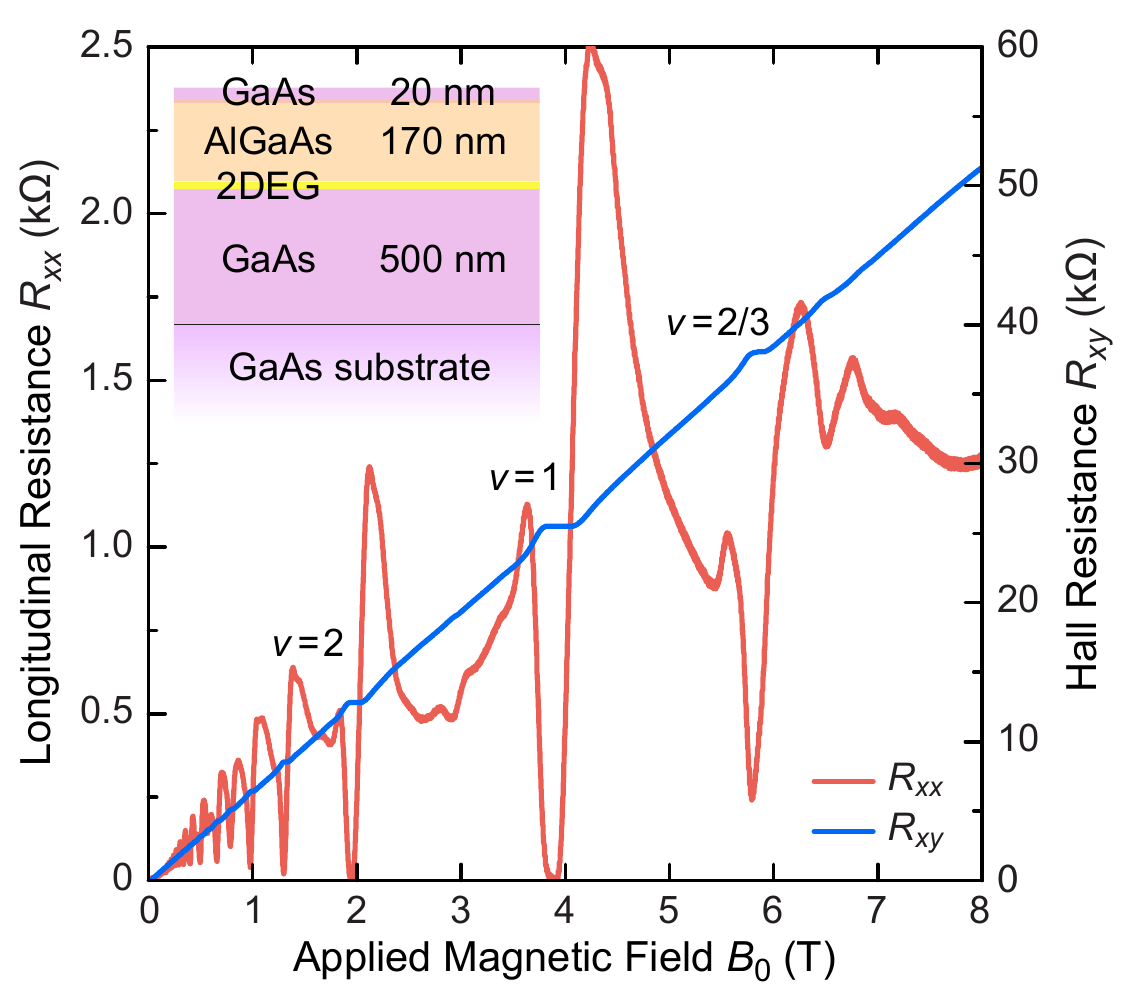}}
\caption{Quantum Hall measurement of the GaAs/AlGaAs 2DEG sample at 0.3~K in dark with zero gate voltage. The inset shows the layer structure of the sample.
\label{efig:qhe}}
\end{figure}

\noindent{\bf 2DEG sample growth and characterization}
Our sample is a single-interface GaAs/Al$_{x}$Ga$_{1-x}$As ($x=0.22$) heterojunction grown by molecular beam epitaxy (MBE) on a 500~$\mu$m thick GaAs wafer. After the growth, the sample is back-polished down to 100~$\mu$m thick in order to enhance the evanescent microwave coupling. The MBE growth consists of a 500~nm thick GaAs layer followed by a 170~nm thick Al$_{x}$Ga$_{1-x}$As ($x=0.22$) spacer and a 20~nm GaAs cap layer to prevent oxidization of the AlGaAs barrier. It is delta-doped with Si doping concentration $1.6\times 10^{12}$~cm$^{-2}$ at a setback of 120~nm above the GaAs/AlGaAs interface containing 2DEG. The 2DEG lies 190~nm below top surface. The electron concentration $n_{0}=0.95\times10^{11}$~cm$^{-2}$ and mobility $\mu=8.6\times10^6$~cm$^2$V$^{-1}$s$^{-1}$ are extracted from our Hall measurement at $T=0.3$~K in dark. In our actual microwave experiment at 0.5~K, the typical zero-gate electron concentration is measured to be about $1\times10^{11}$~cm$^{-2}$. This number is used in our calculation. The uniform magnetic field $B_0$ is supplied by a superconductor coil. In the absence of the holed NdFeB magnet, it can safely reach above 7~T, enabling a quantum-Hall measurement to characterize the sample (see Fig.~\ref{efig:qhe}). When the NdFeB magnet is present, the applied field is limited by practical concerns to at most 0.5~T, beyond which a huge magnetic torque is exerted onto the magnet, risking damage to the sample underneath.

\begin{figure}[ht]
\centerline{\includegraphics[scale=0.68]{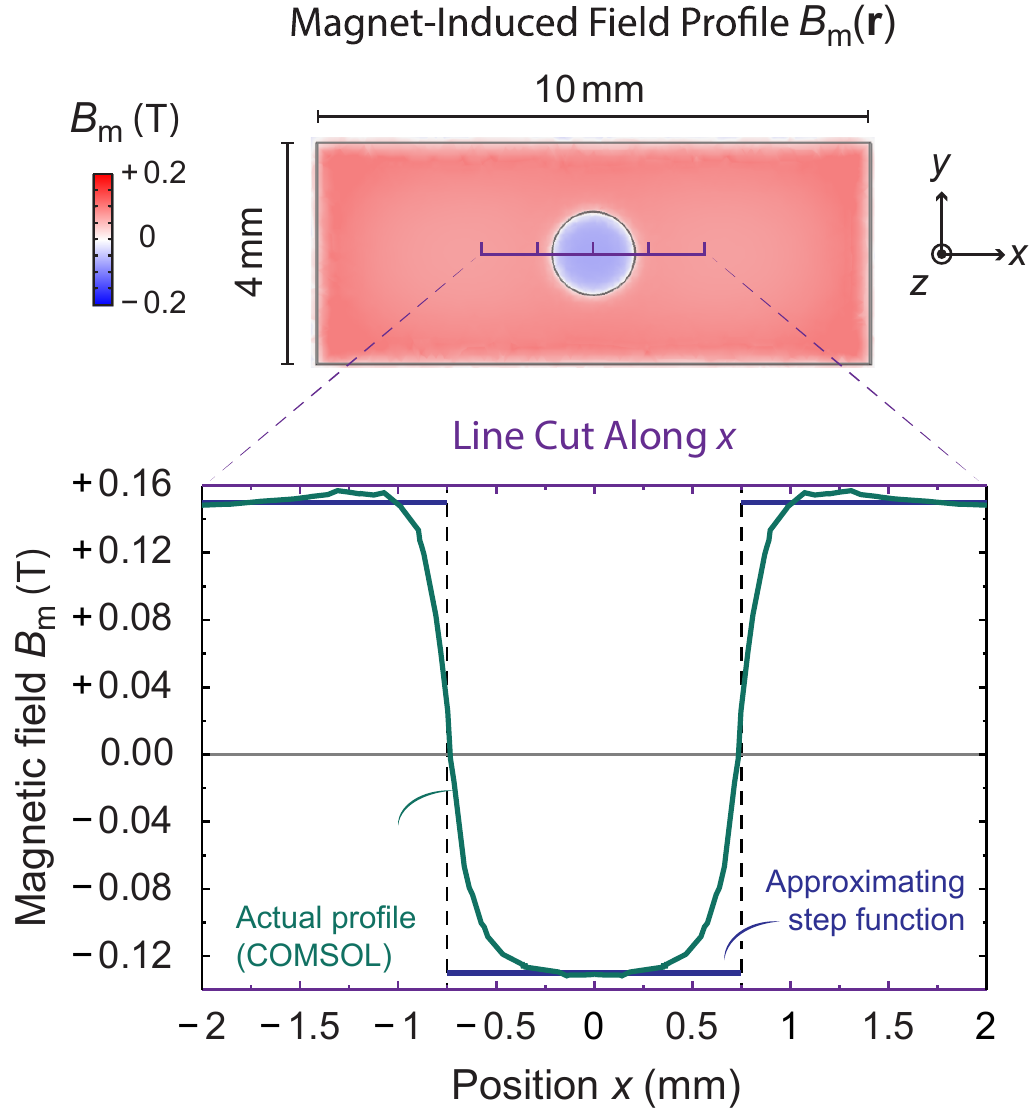}}
\caption{Calculated magnetic field profile of the holed NdFeB magnet at low temperature. The key parameter is imported from the room-temperature Hall-probe measurement and the well-known temperature-dependence in literature.
	\label{efig:magnet}}
\end{figure}

\noindent{\bf NdFeB magnet design and characterization}
The NdFeB magnet is 10~mm long, 4~mm wide, and 1~mm thick, and the hole radius is 0.75~mm. It is produced by sintering NdFeB powders in a custom mold and subsequently magnetizing it along the thickness direction. At room temperature, Hall-probe measurements indicate that the holed magnet provides approximately $\pm0.18$~T remanent magnetic field in the surface area inside and outside the hole. With this value, and taking into account the known anisotropic reduction of the magnetism of NdFeB at cryogenic temperatures \cite{Garcia2000PRL,Strnat1985Proc,TECHNotes}, we are able to simulate out the magnetic field profile over the entire magnet at low temperature (see Fig.~\ref{efig:magnet}) using a finite-element software (Comsol Multiphysics). From the results, we infer that the two key parameters of Eq.~(2) in the main text, namely, a sign-changing field strength $\DBM\approx\pm0.14$~T and a overall shift $\BBM\approx0.01$~T. These are the values used in our theoretical calculations in Fig.~2c and 2d of the main text, demonstrating excellent agreement between theory and experiment with no fitting parameters.

\noindent{\bf Theoretical development and calculation scheme}
The domain-boundary magnetoplasmon (DBMP) modes in our device can be accurately calculated. The evanescent nature of DBMPs and the presence of encapsulating metals, which screen away the long-range part of Coulomb interaction, allow us to focus on the region around and inside the circle $r\lesssim a=0.75$~mm. We can legitimately take a circularly-symmetric model system cut off at a radius $R=10$~mm~$\gg a$, where the scalar potential $\Phi$ is grounded $\Phi(r=R,\varphi)=0$. This rigid boundary condition does not affect the DBMPs far inside.

The associated eigenproblem is most conveniently expressed in the chiral representation~\cite{Jin:2016},
\begin{subequations}
\begin{align}
\omega \JR(r,\varphi) &= + \OmegaC(r) \JR(r,\varphi) \\
 & + \frac{e^2n_0}{\omega_0m_*}\frac{\Ee^{-\Ii\varphi}}{\Ii\sqrt{2}} \left[\partial_r - \frac{\Ii}{r}\partial_\varphi \right] \JD(r,\varphi) , \nonumber\\
\omega \JD(r,\varphi) & = \omega_0\hat{V} \frac{\Ee^{+\Ii\varphi}}{\Ii\sqrt{2}} \left[\partial_r + \frac{\Ii}{r}\partial_\varphi \right] \JR(r,\varphi) \\
 & + \omega_0\hat{V} \frac{\Ee^{-\Ii\varphi}}{\Ii\sqrt{2}} \left[\partial_r - \frac{\Ii}{r}\partial_\varphi \right] \JL(r,\varphi), \nonumber\\
\omega \JL(r,\varphi) &= - \OmegaC(r) \JL(r,\varphi) \\
 & + \frac{e^2n_0}{\omega_0m_*}\frac{\Ee^{+\Ii\varphi}}{\Ii\sqrt{2}} \left[\partial_r + \frac{\Ii}{r}\partial_\varphi \right] \JD(r,\varphi) . \nonumber
\end{align}
\end{subequations}
The basic field components are the right-circulating current $\JR\equiv\frac{1}{\sqrt{2}}(j_{r}-\Ii j_{\varphi})\Ee^{-\Ii\varphi}$, the left-circulating current $\JL\equiv\frac{1}{\sqrt{2}}(j_{r}+\Ii j_{\varphi})\Ee^{+\Ii\varphi}$, and the ``scalar-potential (density-fluctuation)"  current $\JD\equiv\omega_0\Phi$. Here, $\omega_0\equiv\sqrt{ e^2n_0 / m_*R}$ is a characteristic plasmon frequency, $\OmegaC(r) = eB(r)/m_*c$ is the $r$-dependent cyclotron frequency, and $\hat{V}$ is the Coulomb integration operator,
\begin{equation}
\hat{V} \rho(r,\varphi) = \int_0^R r'\Dd r' \int_0^{2\pi} \Dd\varphi'\ V(|\bm{r}-\bm{r}'|) \rho(r',\varphi'),
\end{equation}
with $V(|\bm{r}-\bm{r}'|)$ being the screened Coulomb interaction in real space, which does not have a simple form.

The eigensolutions with a given angular wavenumber $m$ and obeying the hard-wall boundary condition are linear expansion of Bessel functions,
\begin{equation}
	j_{s}(r,\varphi) =\Bigg[ \sum_{n=1}^{N\rightarrow\infty} A_{n,s} \mathrm{J}_{m+s} (q_{m n} r)\Bigg] \Ee^{+\Ii(m+s)\varphi}.
\end{equation}
Here $s=-1,0,+1$ resembles a spin index referring to the $\JR$, $\JD$, $\JL$ components, respectively. $q_{m n}=\zeta_{m n}/R$ are discretized radial wavenumbers in which $\zeta_{m n}$ is the $n$th zero of the $m$th order Bessel function $\mathrm{J}_m(\zeta)$. The expansion is practically cutoff at a large finite $N$ up to a desired spectral resolution. (In our calculations, we use $N=2000$.) With such discretized cylindrical-wave bases, we can rigorously prove that the screened Coulomb interaction relates $\rho$ and $\Phi$ by $\Phi(q_{m n}) = V(q_{m n}) \rho(q_{m n})$, where $V(q_{m n})$ follows Eq.~(3) in the main text.

The matrix-form eigenequation in the cylindrical-wave bases reads
\begin{equation}
	\frac{\omega}{\omega_0}
	\begin{pmatrix}
		\mathbf{A}_{+1} \\
		\mathbf{A}_{0} \\
		\mathbf{A}_{-1}
	\end{pmatrix}
	=
	\begin{pmatrix}
		+ \mathbf{W} &  +\frac{q_{m n}R}{\Ii\sqrt{2}} \mathbf{I} & 0 \\
		- \frac{2\pi\beta(q_{m n})}{\Ii\sqrt{2}} \mathbf{I} & 0 &  + \frac{2\pi\beta(q_{m n})}{\Ii\sqrt{2}} \mathbf{I} \\
		0 &  -\frac{q_{m n}R}{\Ii\sqrt{2}} \mathbf{I} & -  \mathbf{W}
	\end{pmatrix}
	\begin{pmatrix}
		\mathbf{A}_{+1} \\
		\mathbf{A}_{0} \\
		\mathbf{A}_{-1}
	\end{pmatrix}
	.
\end{equation}
Here $\mathbf{A}_s=(\mathcal{A}_{1,s},\mathcal{A}_{2,s},\dots,\mathcal{A}_{N,s})^{\text{T}}$, $\mathbf{I}$ is an $N\times N$ identity matrix, $\mathbf{W}$ is an $N\times N$ full matrix determined by the magnetic-field profile. If $B(r)=B_0$, then $\mathbf{W}=(\OmegaC/\omega_0)\mathbf{I}$ is diagonal with the constant cyclotron frequency $\OmegaC= eB_0 / m_*c$, and the usual bulk MP modes can be recovered~\cite{Jin:2016}.

The generally radially-varying magnetic field in our problem results in scattering between different radial index $n$ (within the $s=-1$ and $+1$ chiral subspace though) and hence localization of new edge modes at the magnetic-domain boundary. We can find the matrix $\mathbf{W}$ via the expansion
\begin{subequations}
\begin{align}
\OmegaC (r) \mathrm{J}_{m-1}(q_{m n}r) \equiv \omega_0 \sum_{n'} W_{nn'}\mathrm{J}_{m-1}(q_{m n'}r),\\
\OmegaC (r) \mathrm{J}_{m+1}(q_{m n}r) \equiv \omega_0 \sum_{n'} W_{nn'}\mathrm{J}_{m+1}(q_{m n'}r),
\end{align}
\end{subequations}
in which $W_{nn'}$ are the elements of $\mathbf{W}$. Upon laborious calculations involving Bessel integrals~\cite{Abramowitz2013Book}, we can get
\begin{align}
\mathbf{W} = \frac{e}{\omega_0m_*c} \left[ 2\DBM \mathbf{Y}^{-1} \mathbf{X} + (B_0+\BBM-\DBM) \mathbf{I} \right] ,
\end{align}
where $\mathbf{X}$ and $\mathbf{Y}$ are $N\times N$ matrices too, whose elements are
\begin{subequations}
\begin{align}
X_{nn'} &=  \int_0^{\tilde{a}} \tilde{r}\Dd\tilde{r}\ \mathrm{J}_{m-1}(\zeta_{m n}\tilde{r}) \mathrm{J}_{m-1}(\zeta_{m n'}\tilde{r}) \\
&=
\begin{cases}
	\!\begin{aligned}
		\DS\frac{\tilde{a}}{\zeta_{m n}^2-\zeta_{m n'}^2}
		\Big\{& \zeta_{m n} \mathrm{J}_{m}(\zeta_{m n}\tilde{a}) \mathrm{J}_{m-1}(\zeta_{m n'}\tilde{a}) \\
		& - \zeta_{m n'} \mathrm{J}_{m}(\zeta_{m n'}\tilde{a}) \mathrm{J}_{m-1}(\zeta_{m n}\tilde{a}) \Big\}
	\end{aligned}
	\quad &\text{for }n\neq n',\\
	\!\begin{aligned}
		\DS\frac{\tilde{a}}{2\zeta_{m n}}
		\Big\{& \tilde{a}\zeta_{m n} \mathrm{J}^2_{m}(\zeta_{m n}\tilde{a}) + \tilde{a}\zeta_{m n} \mathrm{J}^2_{m-1}(\zeta_{m n}\tilde{a}) \\
		&- 2(m-1) \mathrm{J}_{m}(\zeta_{m n}\tilde{a}) \mathrm{J}_{m-1}(\zeta_{m n}\tilde{a}) \Big\}
	\end{aligned}
	\quad &\text{for }n=n',
\end{cases}\nonumber\\
Y_{nn'} &= \delta_{nn'}\frac{1}{2} \mathrm{J}^2_{m-1}(\zeta_{m n}),
\end{align}
\end{subequations}
where $\tilde{a}\equiv a/R$ and $\tilde{r}\equiv r/R$.

\ \vspace{1cm}

\noindent{\bf Author Contributions}
D.J. and Y.X. designed the experiment and fabricated the device. D.J. and T.C. performed the calculation and drafted the manuscript. Y.X., S.W. and K.Y.F. tested the device. D.J. and M.F. carried out the measurement. G.C.G., S.F. and Q.H. grew and processed the MBE samples. Y.W., L.E., M.J.M., N.X.F. and X.Z. provided the guidance and joined the discussion. X.Z. led the project. All authors contributed to the manuscript.

\end{document}